\documentclass[10pt]{article}
\usepackage[T1]{fontenc}
\usepackage{lmodern}
\usepackage[a4paper,margin=0.85in]{geometry}
\usepackage{amsmath,amssymb,amsthm,booktabs,microtype}
\usepackage{listings}
\usepackage{xcolor}
\usepackage{titling}
\usepackage{xurl}
\usepackage{graphicx}
\usepackage[hidelinks]{hyperref}
\hypersetup{pdftitle={A 51-Addition Alternative-Basis Kernel for Rank-23 3x3 Matrix Multiplication},pdfauthor={Joshua Stapleton and A. I. Perminov}}

\newcommand{\vecrow}{\operatorname{vec}_{\rm row}}

\def\keywords#1{\par{{\bfseries Keywords:}\enspace\ignorespaces#1.\par}}

\lstdefinestyle{py}{
    language=Python,
    basicstyle=\ttfamily\small,
    keywordstyle=\color{blue!70!black}\bfseries,
    commentstyle=\color{gray}\itshape,
    stringstyle=\color{purple!70!black},
    showstringspaces=false,
    tabsize=4,
    breaklines=true,
    frame=single,
    captionpos=b
}

\title{A 51-Addition Alternative-Basis Kernel for\\Rank-23 $3\times3$ Matrix Multiplication}
\author{
    \href{https://orcid.org/0009-0008-5257-3827}
    {\includegraphics[scale=0.06]{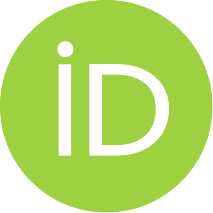}\hspace{1mm}\textbf{Joshua Stapleton}} \\
    Department of Mathematics,\\
    Imperial College\\
    London \\
    \texttt{josh.stapleton23@alumni.imperial.ac.uk}
    \and
    \href{https://orcid.org/0000-0001-8047-0114}
    {\includegraphics[scale=0.06]{orcid.pdf}\hspace{1mm}\textbf{Andrew Perminov}} \\
    Research Center for TAI,\\
    Institute for System Programming\\
    Moscow \\
    \texttt{perminov@ispras.ru}
}
\date{September 23, 2026}
\begin{document}
\maketitle
\begin{abstract}
We give a rank-23 algorithm for $3\times3$ matrix multiplication using
51 additions and subtractions in alternative bases. The input and output
conversions require five further additions, giving 56 additions in
ordinary coordinates. The construction combines
linear-program reduction with a search over sparse basis changes, and the
resulting programs are distributed as a machine-checkable certificate. We
verify correctness by exact coefficient expansion and distinguish the
kernel cost from the cost of a complete multiplication.
\end{abstract}

\keywords{fast matrix multiplication; bilinear algorithms; alternative basis; additive complexity}

\section{Introduction}
Fast matrix multiplication trades scalar multiplications for linear
combinations of matrix entries. The multiplication count therefore
describes only part of the arithmetic work: preparing the input forms
and reconstructing the output also require operations. At a fixed rank,
improving these linear computations is a separate optimisation problem.

The alternative-basis framework~\cite{ks, sparsify} allows these computations to
be performed in coordinates that simplify the kernel. The benefit must
be considered together with the cost of converting between those
coordinates and ordinary matrix entries.

This note presents a rank-23 kernel for $3\times3$ multiplication using
51 additions and subtractions, with five further additions for basis
conversion. This lowers the kernel count of Nielsen's 52-addition
construction~\cite{nielsen} by one. The complete cost is 56 additions,
one more than the ordinary-coordinate construction of Karunaratne and
Idamekorala~\cite{ki55}. We describe the search, give explicit programs,
and verify their correctness.

\section{Result and arithmetic model}
The alternative-basis framework~\cite{ks} separates the cost of evaluating a
bilinear kernel from that of converting input and output coordinates.

Each binary operation $x+y$ or $x-y$ costs one addition. Copies, zero and
permutations are free; left-by-right products are counted separately.
Any standalone unary negations, nonunit scalar multiplications or divisions
must be accounted for explicitly. When comparing with earlier signed-addition
programs, we retain those sources' convention of free unary signs.
Vectorization is row-major.

We write $U,V$ (each $23\times9$) and $W$ ($9\times23$) for the
ordinary-coordinate scheme. For invertible $9\times9$ basis conversions $P,Q,R$,
write $\widetilde U=UP^{-1}$, $\widetilde V=VQ^{-1}$ and
$\widetilde W=R^{-1}W$ for the transformed maps. Thus
\begin{equation}\label{eq:identity}
\vecrow(AB)=R\widetilde W\bigl((\widetilde U P\vecrow(A))
\odot(\widetilde V Q\vecrow(B))\bigr).
\end{equation}
The ordinary-coordinate triple $U,V,W$ satisfies Brent's equations.
The transformed maps $\widetilde U,\widetilde V,\widetilde W$ describe
multiplication in alternative coordinates and generally do not satisfy
those same equations without the conversions.

\paragraph{Result and certificate.}
The displayed kernel programs use $13+12+26=51$ additions and
subtractions. The conversions use $1+3+1=5$, giving a complete
23-product, 56-addition algorithm. All three conversion matrices have
determinant $+1$. Exact expansion of the programs verifies all 729
coefficients in~\eqref{eq:coefficients}. The programs require no
standalone negations, nonunit scalar operations or divisions, and retain
the order of the two factors in each product, so they apply over any
associative ring.

For verification, expand the programs to recover
$U=\widetilde U P$, $V=\widetilde V Q$ and $W=R\widetilde W$.
These matrices must satisfy, with zero-based indices,
\begin{equation}\label{eq:coefficients}
\sum_{t=0}^{22} W_{3i+\ell,t}
U_{t,3i'+j}V_{t,3j'+\ell'}
=\delta_{i,i'}\delta_{j,j'}\delta_{\ell,\ell'}
\end{equation}
for all $i,i',j,j',\ell,\ell'\in\{0,1,2\}$, i.e., all 729
coefficients of the multiplication tensor.

The matrices, basis conversions and straight-line programs are available
as a JSON certificate, together with the Python/NumPy checker
\texttt{verify\_certificate.py}, in the accompanying repository~\cite{certificate}:
\begin{center}
\url{https://github.com/Joshua-Stapleton/51-addition-alt-basis-3x3-rank23}.
\end{center}

\section{Construction}

The construction has two stages: obtaining short linear programs for a
rank-23 scheme, and finding bases that simplify those programs further.
We describe these stages in turn, accounting separately for the kernel
and the conversions needed to use it in ordinary coordinates.

The factorization is derived within the flip-graph framework
\cite{flip_graph_framework}. Starting from this factorization, addition
reduction is performed using a combination of PlinOpt~\cite{plinopt},
fmm\_add\_reduction~\cite{fmm_add_reduction}, and dedicated
algorithms developed within the LEO framework~\cite{leo}. The present work
therefore treats the reducer as a black box and focuses on the
basis-search step that follows it.

Applying the reducer to the flip-graph factorizations yields approximately
$1.6 \times 10^{5}$ candidate factorizations with 55--58 additions, of which
only about 100 are pairwise distinct. The subsequent basis search is
performed on this reduced set of unique programs.

For each program we sample a change-of-basis matrix as follows. We begin
with the $9 \times 9$ identity and replace a random number of its columns
(between 1 and 9) by random signed combinations of a random subset of
rows (between 2 and 9), with coefficients drawn from $\{-1, +1\}$.
Only nonsingular candidates are admissible; the three factors
may use different bases. Each candidate is applied to one of $U$, $V$ or $W^T$,
and the addition reducer is re-run on the transformed factor. Whenever a candidate yields a strictly smaller addition count,
we additionally count the additions required to apply the inverse
change-of-basis matrix.

Two elementary symmetries of the search space under the
free-unary-sign convention are exploited to avoid generating equivalent candidates. First, permuting the columns of the
generated matrix preserves addition complexity, so after
sampling we sort the columns in lexicographic order. Second, negating a
column does not affect the addition count either, so the first coefficient
of every sampled column combination is fixed to $+1$. Both reductions
substantially decrease the number of equivalent matrices considered by
the search, since such matrices carry no additional information.

For historical context, Table~\ref{tab:costs} records the best-known
rank-23 schemes for $3\times3$ matrix multiplication. The first six
entries are constructions in ordinary coordinates, listed in decreasing
order of addition count: a reduced program for Laderman's
scheme~\cite{laderman,fmm_add_reduction}, Stapleton's 60-addition
construction~\cite{stapleton60}, the 59-addition scheme of Mårtensson,
Stankovski Wagner and Stapleton~\cite{martensson59}, Perminov's 58-addition program~\cite{perminov58}, Sun's 56-addition scheme~\cite{sun56}, and the
55-addition construction of Karunaratne and Idamekorala~\cite{ki55}. The
remaining two entries use a change of basis: Nielsen's 52-addition
scheme~\cite{nielsen} and the present work. The columns distinguish
kernel, conversion and complete costs for these specific programs.

\begin{table}[ht]
\centering
\caption{Addition counts for specified rank-23 programs, each using 23
bilinear products. Complete costs include the listed conversions;
Nielsen's conversion count is obtained by replaying the published maps.}
\label{tab:costs}
\begin{tabular}{lrrrrrr}
\toprule
Source & Left & Right & Decoder & Kernel & Conversions & Total\\
\midrule
Laderman (reduced)~\cite{laderman,fmm_add_reduction} & 16 & 16 & 30 & 62 & 0 & 62\\
Stapleton~\cite{stapleton60} & 16 & 16 & 28 & 60 & 0 & 60\\
Mårtensson et al.~\cite{martensson59} & 15 & 15 & 29 & 59 & 0 & 59\\
Perminov~\cite{perminov58} & 16 & 14 & 28 & 58 & 0 & 58\\
Sun~\cite{sun56} & 13 & 13 & 30 & 56 & 0 & 56\\
Karunaratne, Idamekorala~\cite{ki55} & 13 & 14 & 28 & 55 & 0 & 55\\
\noalign{\smallskip}
\hline
\noalign{\smallskip}
Nielsen (alt. basis)~\cite{nielsen} & 12 & 12 & 28 & 52 & 12 & 64\\
\textbf{Our (alt. basis)} & 13 & 12 & 26 & \textbf{51} & 5 & 56\\
\bottomrule
\end{tabular}
\end{table}

\subsection{Complete algorithm}\label{sec:algorithm}
Let $a=\vecrow(A)$ and $b=\vecrow(B)$, where
$a_{3i+j}=A_{ij}$ and $b_{3i+j}=B_{ij}$ for $0\le i,j<3$.
Apply $x=Pa$ and $y=Qb$,
evaluate the displayed forms $u=\widetilde Ux$ and $v=\widetilde Vy$,
and compute $p_j=u_jv_j$ for $j=0,\ldots,22$. Reconstruct
$z=\widetilde Wp$ and return $c=Rz$. Read assignments from left to
right within each row, then from top to bottom; repeated assignments
to $u_{22}$ and $v_{12}$ overwrite their previous values.

The conversions leave all coordinates unchanged except
\[
x_8=a_8-a_6,\qquad
y_2=b_2+b_0,\quad y_6=b_6-b_7,\quad y_8=b_8+b_5,\qquad
c_8=z_8+z_2.
\]
These are the five conversion additions and subtractions.

\begin{scriptsize}
\setlength{\arraycolsep}{4pt}
\[
    U = \left[\begin{array}{rrrrrrrrr}
        1 & 0 & 0 & 0 & 0 & 0 & 0 & 0 & 0 \\
        0 & 1 & 0 & 0 & 0 & 0 & 0 & 0 & 0 \\
        0 & 1 & 0 & 0 & 0 & 0 & 0 & 0 & 0 \\
        0 & 0 & 0 & 1 & 0 & 0 & 0 & 0 & 0 \\
        0 & 0 & 0 & 0 & 1 & 0 & 0 & 0 & 0 \\
        0 & 0 & 0 & 0 & 1 & 0 & 0 & 0 & 0 \\
        0 & 0 & 0 & 0 & 0 & 1 & 0 & 0 & 0 \\
        0 & 0 & 0 & 0 & 0 & 0 & 1 & 0 & 0 \\
        0 & 0 & 0 & 0 & 0 & 0 & 0 & 1 & 0 \\
        0 & 0 & 0 & 0 & 0 & 0 & 0 & 1 & 0 \\
        0 & 0 & 0 & 0 & 0 & 0 & -1 & 0 & 1 \\
        0 & 0 & 0 & 0 & 0 & 1 & -1 & 0 & 1 \\
        0 & 1 & -1 & 0 & 0 & 0 & 0 & 0 & 0 \\
        0 & 1 & -1 & 0 & 0 & 0 & -1 & 0 & 1 \\
        0 & 1 & -1 & 0 & 0 & 0 & 0 & 0 & 1 \\
        0 & 0 & 0 & 1 & 0 & -1 & 1 & 0 & -1 \\
        0 & -1 & 1 & 0 & 0 & 0 & 0 & 1 & -1 \\
        0 & 0 & 0 & 0 & 0 & 1 & 0 & 0 & 1 \\
        0 & 0 & 0 & 0 & 1 & -1 & 0 & 0 & 0 \\
        -1 & 0 & 1 & 1 & 0 & -1 & 1 & 0 & -1 \\
        -1 & 0 & 1 & 1 & 0 & 0 & 1 & 0 & -1 \\
        1 & 0 & -1 & 0 & 0 & 0 & -1 & 0 & 1 \\
        -1 & 0 & 0 & 1 & 0 & 0 & 1 & 0 & 0
    \end{array}\right],
    \quad
    V = \left[\begin{array}{rrrrrrrrr}
        1 & 0 & 1 & 0 & 0 & 0 & 0 & 0 & 0 \\
        0 & 0 & 0 & 1 & 0 & 1 & 1 & 0 & 1 \\
        0 & 1 & 1 & 0 & 1 & 1 & 0 & 1 & 1 \\
        0 & 1 & 0 & 0 & 0 & 0 & 0 & 0 & 0 \\
        0 & 0 & 0 & 1 & 0 & 0 & 0 & 0 & 0 \\
        0 & 0 & 0 & 0 & 1 & 0 & 0 & 0 & 0 \\
        0 & 0 & 0 & 0 & 0 & 0 & 0 & 1 & 0 \\
        -1 & 1 & 0 & 0 & 0 & 0 & -1 & 1 & 0 \\
        0 & 0 & 0 & 1 & 0 & 0 & 0 & 0 & 0 \\
        0 & 0 & 0 & 0 & 1 & 0 & 0 & 0 & 0 \\
        0 & 0 & 0 & 0 & 0 & 0 & 1 & -1 & 0 \\
        1 & 0 & 0 & 0 & 0 & 0 & 1 & -1 & 0 \\
        0 & 0 & 0 & 0 & 0 & 0 & -1 & 0 & -1 \\
        0 & 1 & 1 & 0 & 0 & 0 & -1 & 1 & 0 \\
        0 & 1 & 1 & 0 & 0 & 1 & 0 & 1 & 1 \\
        1 & 0 & 0 & 0 & 0 & 0 & 0 & 0 & 0 \\
        0 & 0 & 0 & 0 & 0 & 1 & 0 & 0 & 0 \\
        1 & 0 & 0 & 0 & 0 & 0 & 1 & 0 & 0 \\
        0 & 0 & 0 & 0 & 0 & 1 & 0 & 0 & 0 \\
        0 & 0 & 0 & 0 & 0 & 1 & 0 & 0 & 1 \\
        0 & 1 & 1 & 0 & 0 & 1 & 0 & 0 & 1 \\
        0 & 1 & 1 & 0 & 0 & 0 & 0 & 0 & 0 \\
        0 & 0 & -1 & 0 & 0 & -1 & 0 & 0 & -1
    \end{array}\right],
\]

\[
    W = \left[\begin{array}{rrrrrrrrrrrrrrrrrrrrrrr}
        1 & 1 & 0 & 1 & 0 & 0 & -1 & 1 & 0 & 0 & 1 & -1 & 0 & 1 & -1 & 0 & 0 & 1 & 0 & 0 & -1 & -1 & -1 \\
0 & 0 & 1 & 1 & 0 & 0 & -1 & 1 & 0 & 0 & 0 & -1 & 0 & 0 & -1 & 0 & 0 & 1 & 0 & 0 & -1 & 0 & -1 \\
0 & 0 & 0 & -1 & 0 & 0 & 1 & -1 & 0 & 0 & -1 & 1 & 1 & -1 & 1 & 0 & 0 & -1 & 0 & 0 & 1 & 1 & 1 \\
0 & 0 & 0 & 0 & 1 & 0 & 1 & 0 & 0 & 0 & -1 & 1 & 0 & 0 & 0 & 1 & 0 & 0 & 0 & 0 & 0 & 0 & 0 \\
0 & 0 & 0 & 1 & 0 & 1 & 1 & 0 & 0 & 0 & 0 & 0 & 0 & 0 & 0 & 0 & 0 & 0 & 0 & 0 & 0 & 0 & 0 \\
0 & 0 & 0 & -1 & 0 & 0 & 0 & 0 & 0 & 0 & 0 & 0 & 0 & 0 & 0 & 0 & 0 & 0 & 1 & -1 & 1 & 1 & 0 \\
0 & 0 & 0 & 0 & 0 & 0 & -1 & 0 & 1 & 0 & 1 & -1 & 0 & 0 & 0 & 0 & 0 & 1 & 0 & 0 & 0 & 0 & 0 \\
0 & 0 & 0 & 0 & 0 & 0 & -1 & 1 & 0 & 1 & 0 & -1 & 0 & 0 & 0 & 0 & 0 & 1 & 0 & 0 & 0 & 0 & 0 \\
0 & 0 & 0 & 0 & 0 & 0 & 1 & -1 & 0 & 0 & -1 & 1 & 1 & -1 & 1 & 0 & 1 & -1 & 0 & 0 & 0 & 0 & 0 \\
    \end{array}\right].
\]

\[
    P = \left[\begin{array}{rrrrrrrrr}
        1 & 0 & 0 & 0 & 0 & 0 & 0 & 0 & 0 \\
        0 & 1 & 0 & 0 & 0 & 0 & 0 & 0 & 0 \\
        0 & 0 & 1 & 0 & 0 & 0 & 0 & 0 & 0 \\
        0 & 0 & 0 & 1 & 0 & 0 & 0 & 0 & 0 \\
        0 & 0 & 0 & 0 & 1 & 0 & 0 & 0 & 0 \\
        0 & 0 & 0 & 0 & 0 & 1 & 0 & 0 & 0 \\
        0 & 0 & 0 & 0 & 0 & 0 & 1 & 0 & 0 \\
        0 & 0 & 0 & 0 & 0 & 0 & 0 & 1 & 0 \\
        0 & 0 & 0 & 0 & 0 & 0 & -1 & 0 & 1 
    \end{array}\right],
    \quad
    Q = \left[\begin{array}{rrrrrrrrr}
        1 & 0 & 0 & 0 & 0 & 0 & 0 & 0 & 0 \\
        0 & 1 & 0 & 0 & 0 & 0 & 0 & 0 & 0 \\
        1 & 0 & 1 & 0 & 0 & 0 & 0 & 0 & 0 \\
        0 & 0 & 0 & 1 & 0 & 0 & 0 & 0 & 0 \\
        0 & 0 & 0 & 0 & 1 & 0 & 0 & 0 & 0 \\
        0 & 0 & 0 & 0 & 0 & 1 & 0 & 0 & 0 \\
        0 & 0 & 0 & 0 & 0 & 0 & 1 & -1 & 0 \\
        0 & 0 & 0 & 0 & 0 & 0 & 0 & 1 & 0 \\
        0 & 0 & 0 & 0 & 0 & 1 & 0 & 0 & 1
    \end{array}\right],
    \quad
    R = \left[\begin{array}{rrrrrrrrr}
        1 & 0 & 0 & 0 & 0 & 0 & 0 & 0 & 0 \\
        0 & 1 & 0 & 0 & 0 & 0 & 0 & 0 & 0 \\
        0 & 0 & 1 & 0 & 0 & 0 & 0 & 0 & 0 \\
        0 & 0 & 0 & 1 & 0 & 0 & 0 & 0 & 0 \\
        0 & 0 & 0 & 0 & 1 & 0 & 0 & 0 & 0 \\
        0 & 0 & 0 & 0 & 0 & 1 & 0 & 0 & 0 \\
        0 & 0 & 0 & 0 & 0 & 0 & 1 & 0 & 0 \\
        0 & 0 & 0 & 0 & 0 & 0 & 0 & 1 & 0 \\
        0 & 0 & 1 & 0 & 0 & 0 & 0 & 0 & 1
    \end{array}\right].
\]
\end{scriptsize}

The computation of $u$ requires 13 additions:
\begin{align*}
    u_{0} = x_{0} &&
    u_{1} = x_{1} &&
    u_{2} = x_{1} &&
    u_{3} = x_{3} &&
    u_{4} = x_{4} &&
    u_{5} = x_{4} \\
    u_{6} = x_{5} &&
    u_{7} = x_{6} &&
    u_{8} = x_{7} &&
    u_{9} = x_{7} &&
    u_{10} = x_{8} &&
    u_{11} = x_{5} + x_{8} \\
    u_{12} = x_{1} - x_{2} &&
    u_{13} = x_{8} + u_{12} &&
    u_{14} = x_{6} + u_{13} &&
    u_{15} = x_{3} - u_{11} &&
    u_{16} = x_{7} - u_{14} &&
    u_{17} = x_{6} + u_{11} \\
    u_{18} = x_{4} - x_{5} &&
    u_{22} = x_{0} - u_{15} &&
    u_{19} = x_{2} - u_{22} &&
    u_{20} = x_{5} + u_{19} &&
    u_{21} = x_{3} - u_{20} &&
    u_{22} = u_{17} - u_{22}
\end{align*}

The computation of $v$ requires 12 additions:
\begin{align*}
    v_{0} = y_{2} &&
    v_{11} = y_{0} + y_{6} &&
    v_{7} = y_{1} - v_{11} &&
    v_{13} = y_{2} + v_{7} &&
    v_{21} = y_{6} + v_{13} &&
    v_{20} = y_{8} + v_{21} \\
    v_{14} = y_{7} + v_{20} &&
    v_{2} = y_{4} + v_{14} &&
    v_{17} = y_{7} + v_{11} &&
    v_{22} = y_{1} - v_{20} &&
    v_{12} = v_{13} - v_{14} &&
    v_{1} = y_{3} - v_{12} \\
    v_{12} = y_{5} + v_{12} &&
    v_{3} = y_{1} &&
    v_{4} = y_{3} &&
    v_{5} = y_{4} &&
    v_{6} = y_{7} &&
    v_{8} = y_{3} \\
    v_{9} = y_{4} &&
    v_{10} = y_{6} &&
    v_{15} = y_{0} &&
    v_{16} = y_{5} &&
    v_{18} = y_{5} &&
    v_{19} = y_{8}
\end{align*}

Next compute the 23 ordered products $p_j=u_jv_j$ for $j=0,\ldots,22$.

And the computation of $z$ requires 26 additional additions:
\begin{align*}
    t_{0} = p_{11} - p_{10} + p_{6} &&
    t_{1} = p_{20} + p_{21} - p_{3} &&
    t_{2} = p_{22} + t_{1} &&
    t_{3} = p_{17} - t_{0} \\
    t_{4} = p_{7} + t_{3} &&
    t_{5} = t_{4} - p_{14} - t_{2} &&
    t_{6} = p_{13} + t_{5} \\
    \\
    z_{0} = p_{0} + p_{1} + t_{6} &&
    z_{1} = p_{2} - p_{10} + p_{21} + t_{5} &&
    z_{2} = p_{12} - t_{6} \\
    z_{3} = p_{4} + p_{15} + t_{0} &&
    z_{4} = p_{5} + p_{6} + p_{3} &&
    z_{5} = p_{18} - p_{19} + t_{1} \\
    z_{6} = p_{8} + t_{3} &&
    z_{7} = p_{9} - p_{10} + t_{4} &&
    z_{8} = p_{16} - t_{2}    
\end{align*}

Finally, apply $c=Rz$ and set $C_{ij}=c_{3i+j}$ for $0\le i,j<3$;
then $C=AB$.

\section{Conclusion}
We have presented a rank-23 algorithm for $3\times3$ matrix multiplication
requiring 51 additions in alternative bases and five further additions for
the input and output conversions. This reduces the kernel count of Nielsen's
construction by one. Including conversions, the complete cost is 56 additions,
one more than the ordinary-basis scheme of Karunaratne and Idamekorala. All
three basis conversions have determinant $+1$, and the programs contain no
standalone negations, nonunit scalar multiplications or divisions, so the
algorithm applies over any associative ring. The scheme is distributed as a
machine-checkable certificate together with a short verification script.

Beyond the specific counts, the construction illustrates a general point:
the arithmetic cost of a bilinear algorithm splits into a kernel part and a
conversion part, and these two quantities can be optimized somewhat
independently. How this balance affects running time in recursive implementations
and how far the kernel count can be pushed below 51 by further basis search
remain open questions. We leave a systematic treatment
of the reduction step and a broader exploration of the trade-off to future
work.

\paragraph{Acknowledgment.}
AI-assisted tools were used for code development, experiment orchestration, and review of this note.

\appendix
\section{Verification Script}
\label{app:verification}

\begin{lstlisting}[style=py,escapeinside={(*@}{@*)},caption={Python script for verification of the 51-addition alt. basis scheme}]
import json
import numpy as np


def multiply3x3_cr51(a: np.ndarray, b: np.ndarray) -> np.ndarray:
    a0, a1, a2, a3, a4, a5, a6, a7, a8 = a.reshape(9)
    b0, b1, b2, b3, b4, b5, b6, b7, b8 = b.reshape(9)

    x0, x1, x2 = a0, a1, a2
    x3, x4, x5 = a3, a4, a5
    x6, x7, x8 = a6, a7, a8 - a6

    y0, y1, y2 = b0, b1, b2 + b0
    y3, y4, y5 = b3, b4, b5
    y6, y7, y8 = b6 - b7, b7, b8 + b5

    # 13 additions
    u0 = x0
    u1 = x1
    u2 = x1
    u3 = x3
    u4 = x4
    u5 = x4
    u6 = x5
    u7 = x6
    u8 = x7
    u9 = x7
    u10 = x8
    u11 = x5 + x8
    u12 = x1 - x2
    u13 = x8 + u12
    u14 = x6 + u13
    u15 = x3 - u11
    u16 = x7 - u14
    u17 = x6 + u11
    u18 = x4 - x5
    u22 = x0 - u15
    u19 = x2 - u22
    u20 = x5 + u19
    u21 = x3 - u20
    u22 = u17 - u22

    # 12 additions
    v0 = y2
    v11 = y0 + y6
    v7 = y1 - v11
    v13 = y2 + v7
    v21 = y6 + v13
    v20 = y8 + v21
    v14 = y7 + v20
    v2 = y4 + v14
    v17 = y7 + v11
    v22 = y1 - v20
    v12 = v13 - v14
    v1 = y3 - v12
    v12 = y5 + v12
    v3 = y1
    v4 = y3
    v5 = y4
    v6 = y7
    v8 = y3
    v9 = y4
    v10 = y6
    v15 = y0
    v16 = y5
    v18 = y5
    v19 = y8

    p0 = u0 * v0
    p1 = u1 * v1
    p2 = u2 * v2
    p3 = u3 * v3
    p4 = u4 * v4
    p5 = u5 * v5
    p6 = u6 * v6
    p7 = u7 * v7
    p8 = u8 * v8
    p9 = u9 * v9
    p10 = u10 * v10
    p11 = u11 * v11
    p12 = u12 * v12
    p13 = u13 * v13
    p14 = u14 * v14
    p15 = u15 * v15
    p16 = u16 * v16
    p17 = u17 * v17
    p18 = u18 * v18
    p19 = u19 * v19
    p20 = u20 * v20
    p21 = u21 * v21
    p22 = u22 * v22

    # 26 additions
    t0 = p11 - p10 + p6
    t1 = p20 + p21 - p3
    t2 = p22 + t1
    t3 = p17 - t0
    t4 = p7 + t3
    t5 = t4 - p14 - t2
    t6 = p13 + t5

    z0 = p0 + p1 + t6
    z1 = p2 - p10 + p21 + t5
    z2 = p12 - t6
    z3 = p4 + p15 + t0
    z4 = p5 + p6 + p3
    z5 = p18 - p19 + t1
    z6 = p8 + t3
    z7 = p9 - p10 + t4
    z8 = p16 - t2

    c = np.array([z0, z1, z2, z3, z4, z5, z6, z7, z8 + z2])
    return c.reshape(3, 3)


def main():
    with open("3x3x3_m23_cr51_alt_basis_certificate.json", "r") as f:
        data = json.load(f)

    U = np.array(data["original_matrices"]["U"])
    V = np.array(data["original_matrices"]["V"])
    W = np.array(data["original_matrices"]["W"])

    Uc = np.array(data["changed_matrices"]["U"])
    Vc = np.array(data["changed_matrices"]["V"])
    Wc = np.array(data["changed_matrices"]["W"])

    P = np.array(data["basis"]["P"])
    Q = np.array(data["basis"]["Q"])
    R = np.array(data["basis"]["R"])

    # check valid basis conversion
    assert np.allclose(U, Uc @ P)
    assert np.allclose(V, Vc @ Q)
    assert np.allclose(W, R @ Wc)

    # check Brent equations
    tensor = np.zeros((9, 9, 9), dtype=np.int32)
    for i in range(3):
        for j in range(3):
            for k in range(3):
                tensor[i * 3 + k, k * 3 + j, i * 3 + j] = 1

    assert np.allclose(tensor, np.einsum("ri,rj,kr->ijk", U, V, W))

    for _ in range(1000):
        a, b = np.random.randn(3, 3), np.random.randn(3, 3)
        c = a @ b

        x = P @ a.reshape(9)
        y = Q @ b.reshape(9)

        u = Uc @ x
        v = Vc @ y
        p = u * v
        z = Wc @ p

        # check the full scheme in the new basis
        assert np.allclose(c, (R @ z).reshape(3, 3))
        # check the reduced scheme in the new basis
        assert np.allclose(c, multiply3x3_cr51(a, b))

    print("All tests passed")


if __name__ == '__main__':
    main()
\end{lstlisting}

\end{document}